\documentclass[a4paper]{jpconf}
\usepackage{graphicx}
\usepackage{amssymb, amsmath}
\usepackage{multirow}
\usepackage[section]{placeins}
\usepackage{float}
\begin{document}

\title{Uncertainties of optical-model parameters for the study of the threshold anomaly}

\author{Daniel Abriola$^1$, A. Arazi$^{1,}$$^2$, J. Testoni$^1$ , F. Gollan$^2$ and G.V. Mart\'\i$^1$,}

\address{$^1$ Laboratorio TANDAR, Comisi\'on Nacional de Energ\'\i a At\'omica,
Avda. General Paz 1499, (B1650KNA) San Mart\'\i n, Argentina}
\address{$^2$ Consejo Nacional de Investigaciones Cient\'\i ficas y T\'ecnicas,
Avda. Rivadavia 1917, (C1033AAJ) Buenos Aires, Argentina}

\ead{abriola@tandar.cnea.gov.ar}

\begin{abstract}
In the analysis of elastic-scattering experimental data, optical-model parameters (usually, depths of real and imaginary potentials) are fitted and conclusions are drawn analyzing  their variations at bombardment energies close to the Coulomb barrier (threshold anomaly). The judgement about the shape of this variation (related to the physical processes producing this anomaly) depends on these fitted values but the robustness of the conclusions strongly depends  on the uncertainties with which these parameters are derived. We will show that previous published studies have not used a common criterium for the evaluation of the parameter uncertainties. In this work, a study of these  uncertainties is presented, using conventional statistic tools as well as  bootstrapping techniques. As case studies, these procedures are applied to re-analyze detailed elastic-scattering data for the $^{12}$C + $^{208}$Pb and the $^6$Li + $^{80}$Se systems.
\end{abstract}
\section{Introduction}
In the study of elastic scattering of atomic nuclei at low energies there has been a long-lasting interest in the so-called threshold anomaly (TA \cite{sat91}). For the case of tightly bound nuclei, this phenomenon, related to the closure of reaction channels, consists in a decrease of the depth of imaginary part of the optical potential at  bombarding energies below the Coulomb barrier $V_{B}$. Due to the causality-related dispersion relation (DR \cite{sat91}) linking the real and imaginary parts of optical potentials, the depth of the real part also varies strongly, peaking around $V_{B}$ (see Refs.~\cite{abr89, abr92, sat91}). In the case of weakly bound projectiles, the coupling to  nonelastic channels (e.g. breakup) generates a repulsive polarization
potential~\cite{sak87} that can produce either the absence of TA ({\em no TA}~\cite{mah86, kee94,gom04,fig06}) or the  so-called breakup threshold anomaly (BTA~\cite{gom04,hus06,fig06}), in which the imaginary potential increases while approaching $V_{B}$ and, conversely, there is a reduction on the real part of the potential~\cite{hus06,can06}.

For an unambiguous determination of the kind of anomaly, the variation of the optical-model parameter values as a function of the energy must be evaluated taking into account their uncertainties (hereafter called {\em parameter uncertainties}), which must be derived from  the experimental angular distribution data and their own uncertainties (hereafter called {\em experimental uncertainties}). However, publications on the subject apply different, sometimes non-rigorous criteria for the estimation of the parameter uncertainties or do not consider uncertainties at all. Here, a study of the parameter uncertainties is presented, using conventional statistic tools as well as the bootstrap technique~\cite{pre07}. It is hoped that these methods will help to achieve a better determination of the parameters of interest in the characterization of the TA.
In section \ref{SectionTheory} we discuss mathematical and physical issues that influence the determination of uncertainties when adjusting optical-model parameters using experimental angular distributions, in section \ref{SectionCaseStudies} the $^{12}$C + $^{208}$Pb and the $^6$Li + $^{80}$Se systems will be reanalyzed and their uncertainties evaluated with different techniques. Finally, in section \ref{SectionConclusions} we make our  final remarks.

\section{The determination of optical-model parameter uncertainties } \label{SectionTheory}
In this section we will firstly consider the $\chi^2$ test (\ref{sec-chi2}) since it is the most popular method to evaluate the optical-model parameter uncertainties, we will review the different $\chi^2$ criteria
found in the literature (\ref{criteria}) and finally we will consider the two techniques that we will apply to the case studies in the next section, namely: {\em covariance ellipses} (\ref{errorellipses}) and {\em bootstrap method} (\ref{bootstrap}).
\subsection{The $\chi^2$ test}\label{sec-chi2}
Let us consider a series of experimental pairs $x_i, y_i(x_i)$, where the $y_i$ values have an uncertainty  $\Delta y_i$ (statistical, background substraction, etc.) which is assumed to be normally distributed.
To compare these experimental data with a theoretical model that adjust the data with a function $y=f(x)$ taking into account the experimental uncertainties, we could use an adjusting program that minimizes the value of $\chi^2$ defined as
\begin{equation}\label{definitionchi2}
\displaystyle{ \chi^2 = \sum_i^N  R_i^2  = \sum_i^N \biggl[\frac{y_i - f(x_i)}{\Delta y_i}\biggr]^2\,},
\end{equation}
where $N$ is the number of experimental points and $R_i$ are the so-called  {\em residuals}.
The adjusting program should minimize the effective distance between the experimental results $y_i$ and their theoretical estimation $f(x_i)$.
The simplest case to study is the repetition of a measurement of a physical quantity (say a life-time or the speed of light)
by several laboratory groups. The theoretical model $f(x_i)$ could be as simple as the construction of the best evaluated value of that physical quantity, e.g. the arithmetic or weighted average. In this simple case the expected $\chi^2$ should be close to $N-1$, where $N$ is the number of measurements. If not,  procedures such as the proposed by Birge \cite{bir32,kac08, kes11} should be applied. In its original form, it applies for interlaboratory evaluations (external uncertainty) when $\chi^2/(N-1) > 1$ and consists in multiplying the experimental (internal) uncertainties by  $\sqrt{\chi^2/(N-1)}$. With the increased uncertainties a new, normalized value $\chi^2 {'}$ is obtained, here using $N$ instead of $N-1$ for large number of points:
\begin{equation}\label{normchi2}
{\chi^2{'} = \displaystyle{\sum_{i=1}^N \frac{[y_i - f(x_i)]^2} {\Delta Y_i^2}} =
\sum_{i=1}^N \frac{[y_{i} - f(x_{i})]^2}{(\sqrt{\chi^2/N} \Delta y_i)^2} =
\sum_{i=1}^N \frac{[y_i - f(x_i)]^2}{\Delta y_i^2} (\frac{N}{\chi^2})= \chi^2 (\frac{N}{\chi^2})=N}
\end{equation}
We consider now cases in which the theoretical function $f$ has $k$ parameters $\boldsymbol{\alpha} = (\alpha_1,...,\alpha_k)$, denoted from now on as $f(x_i|\boldsymbol{\alpha})$, that should be adjusted to achieve  the best fit of $f$ to experimental data. It is a well-known rule \cite{bev03,pre07} that the uncertainty with which a single, uncorrelated parameter $\alpha_j$ is adjusted, can be obtained varying this parameter around its optimum value $\alpha_{j0}$ (while all others parameters are optimized) until the value of $\chi^2$ increments in one unit, i.e.:
\begin{equation}\label{chi2+1}
\chi^2=\chi^2_0+1.
\end{equation}
Here $\chi^2_{0}$ denotes the minimum value of $\chi^2$. For this rule to be valid the following conditions must apply:
\begin{nopagebreak}[4]
\begin{itemize}
\item The uncertainties of raw experimental data $\Delta y_i$ are properly determined.
\item There are no significant systematic errors.
\item The theoretical model is a true description of the data being studied.
\end{itemize}
\end{nopagebreak}
 As consequence of the previous conditions  the residuals $R_i$ will follow a normal  distribution with mean value $\langle R_i \rangle =0$ and variance $\sigma^2_R=1$ (we will denote this $R_i \sim {\cal N}(0,1)$) which  implies $\chi^2_0 \sim \nu $ (although the reciprocal is not necessarily true). Here $\nu$ is the number of degrees of freedom $\nu = N - n_f$ being $n_f$ the number of free parameters. In most of the cases $n_f = 2$ and $N \geq 20$, thus $\nu \approx N$ and no significant difference arises from the use of N instead of $\nu$.

As we will see in section \ref{criteria} many data sets analyzed in the area of low-energy elastic scattering do not fulfil all the conditions previously mentioned, but the rule of eq. \ref{chi2+1} is nevertheless applied. Hence the uncertainty of the obtained parameters is underestimated. For these cases, we will extend the procedure of Birge (eq. \ref{normchi2}) for the experimental adjustment of one or several parameters $\boldsymbol{\alpha}$ of a theoretical function $f(x_i|\boldsymbol{\alpha})$ in the following way:
\begin{equation}\label{birge-parameter}
{\chi^2{'} = \displaystyle{\sum_{i=1}^N \frac{[y_i - f(x_i|\boldsymbol{\alpha})]^2} {(\sqrt{\chi^2_0/\nu} \Delta y_i)^2}} =
 \frac{\chi^2} {(\chi^2_0/\nu)}}
\end{equation}
For the optimum set of parameters $\boldsymbol{\alpha_0}$ we obtain
\begin{equation}\label{birge-parameter_0}
{\chi^2_0{'} = \frac{\chi^2_0} {(\chi^2_0/\nu)}}.
\end{equation}
Now, with the experimental uncertainties $\Delta y_i$ scaled by the Birge factor $\sqrt{\chi^2_0/\nu}$ the rule of eq. \ref{chi2+1} can be applied:
\begin{equation}\label{chi2'}
\chi^2{'}=\chi^2_0{'}+1.
\end{equation}
This does not imply that the experimental uncertainties $\Delta y_i$ are effectively changed in fact the original ones should be reported. To avoid confusions we prefer to replace in eq. \ref{chi2'} with eq. \ref{birge-parameter} and \ref{birge-parameter_0} to obtain the equivalent rule
\begin{equation}\label{chi2+chi20/nu}
\chi^2=\chi^2_0+\chi^2_0/\nu\,.
\end{equation}
In the case of adjusting several (even correlated) parameters the rule of eq. \ref{chi2} should be extended as $\chi^2=\chi^2_0+\Delta \chi^2$, where for two parameters and a 68\% confidence level $\Delta \chi^2 = 2.3$. See Ref.~\cite{pre07} (p.~815) for a table of $\Delta \chi^2$ for different confidence levels  and number of adjustable parameters.
In the more general case in which $\chi^2_0/\nu > 1$ and several parameters are adjusted simultaneously eq. \ref{chi2+chi20/nu} is extended to
\begin{equation}\label{chi2+chi20/nudeltachi}
\chi^2=\chi^2_0+\Delta \chi^2{'},~~~~~~\text{where}~~~~~~\Delta \chi^2 {'}=\Delta \chi^2\,\chi^2_0/\nu\,.
\end{equation}
In the study of scattering angular distributions, the experimental data $(x_i,y_i)$ consist in angles  and differential cross-sections $(\theta_i, d\sigma_i/d\Omega)$. The theoretical model $f(x_i|\boldsymbol{\alpha})$ can be either the calculated $d\sigma(\theta_i)/d\Omega $ using the phenomenological Woods-Saxon optical model or the microscopic models involving folding of nuclear densities with appropriated nucleon-nucleon potentials. In this work the S\~{a}o Paulo global microscopic potential (SPP ~\cite{cha97, alv03}) is used with $\alpha_1 = N_R$ and $\alpha_2 = N_I$ (depth of real and imaginary parts of the potential, respectively).

\subsection{Criteria used in previous works }\label{criteria}
\begin{figure}
\includegraphics[width=40pc]{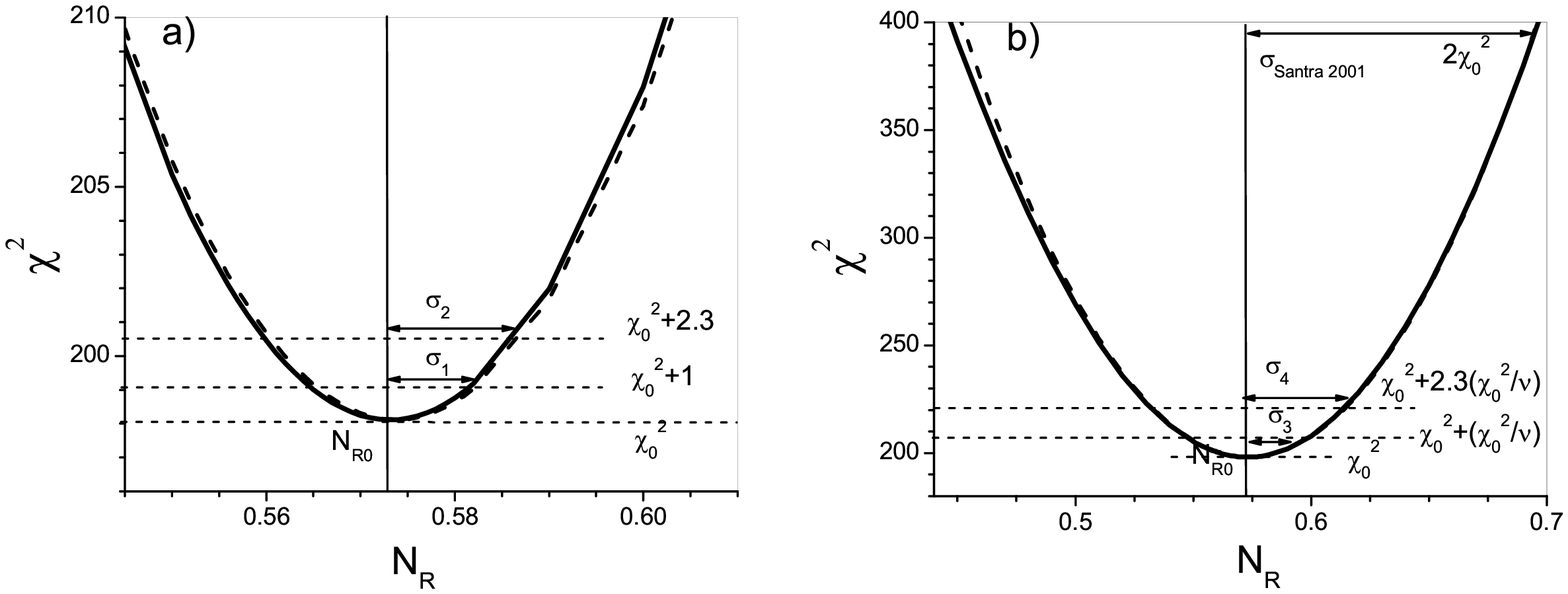}
\vspace{-7cm}
\caption{\label{chi2}$\chi^2$ as a function of $N_R$ for  $^{12}$C + $^{208}$Pb at 84.9 MeV (Plots a) and b) show different scales of the  axes).  Different criteria for the uncertainty interval of $N_R$ are shown: $\sigma_1 \approx 0.0086$, $\sigma_2 \approx  0.013$,   $\sigma_3 \approx  0.0271$, $\sigma_4 \approx  0.041$. The criterium applied in Ref.~\cite{san01} gives $\sigma_{\text{Santra 2001}}\approx  0.121$. }
\end{figure}

Table 1 presents a non-exhaustive list  of experiments studying the TA, either with  weakly or tightly bound projectiles. It can be seen that most studies use the standard rule $\chi^2 \leq  \chi^2_{0}+1$ (eq. \ref{chi2+1}) to determine the uncertainties regardless of their $\chi^2/\nu$ value . We will show in the next section that due to parameter correlations, this rule gives  too small uncertainties bars, even in the best case when  $\chi_0^2/\nu \approx 1$, being even worse when $\chi^2/\nu \gg 1$. It is apparent that this  problem has been recognized but not clearly acknowledged and, thus, many of the different criteria listed in Table 1 attempted to increase the $\chi^2 \leq \chi^2_{0}+1$ limits. A good example of this can be seen in Ref. ~\cite{san01}. Since their angular distribution data for the $^{12}$C + $^{208}$Pb system have an average value  $\langle\chi^2_0/\nu\rangle = 6$, the usual criterium $\chi^2 \leq  \chi_0^2 +1$
would  have clearly been an underestimation of the parameter uncertainties and an extreme criterium  $\chi^2 \leq 2 \chi_0^2$ was used, which, as we will show in section \ref{12Csubsect}, is an overestimation. In Fig.~\ref{chi2} we show the $\chi^2$ vs. $N_R$ curve corresponding to our reanalysis using SPP of the 84.9 MeV angular distribution of Ref.~\cite{san01}. The optimum fit is obtained applying a combination of grid, and steepest descent searching routines and gives:
  $N_{R0}=0.5738$, $N_{I0}=0.5173$, $\chi^2_0 = 198$, $\chi^2_0/\nu = 9.9$. The covariance matrix, including the variance of each of the two parameters $N_R$ and $N_I$ and their covariance,  is calculated using second derivatives calculated numerically around the minimum~\cite{bev03}. The parabolic approximation of the curve describing $\chi^2$ as a function  of $N_R$ is shown in
  Fig.~\ref{chi2} with  dashed lines. Its expression is  given by
\begin{equation}\label{parabolicaprox}
  {\displaystyle{\chi^2 = \frac{1}{\sigma_1^2} ( N_R-N_{R0})^2+ \chi^2_{0} } },\\
\end{equation}
where $\sigma_1^2$ is the variance of $N_R$. This variance determines the value of  $|N_R - N_{R0}|$ that produces a change in one unit in the $\chi^2$, as in eq. \ref{sigma1}. The solid lines in Fig.~\ref{chi2} represent the values corresponding to a grid on $N_R$ while $N_I$ is optimized. It is seen that the parabolic approximation is quite good even at several $\sigma_1$ away from the minimum. Uncertainty limits corresponding to different criteria found in the literature are:
\begin{subequations}\label{criteriossigma}
\begin{align}
    \sigma_1 &\rightarrow  \chi^2 \leq \chi^2_0 +1 \label{sigma1}\\
    \sigma_2 &\rightarrow  \chi^2 \leq \chi^2_{0}+2.3 \label{sigma2} \\
    \sigma_3 &\rightarrow  \chi^2 \leq \chi^2_0 +\chi^2_0/\nu \label{sigma3} \\
    \sigma_4 &\rightarrow  \chi^2 \leq \chi^2_0+2.3\chi^2_0/\nu \label{sigma4}.
\end{align}
\end{subequations}
These equations are not meant to represent different confidence limits (CL) but different criteria. In the particular example of Fig.~\ref{chi2}, $\sigma_3$ represents the $68\%$ CL for the adjustment of the single parameter $N_R$, considered as uncorrelated with $N_I$, while $\sigma_4$ represents the $68\%$ CL for the simultaneous adjustment of the correlated $N_R$ and $N_I$. In this case, since $\chi^2_0/\nu = 9.9$,  $\sigma_1$  and $\sigma_2$ do not represent $68\%$ CL.
In the parabolic approximation of eq. \ref{parabolicaprox} it holds
\begin{equation}\label{receta}
\sigma_2 = \sqrt{2.3}\,\sigma_1,~~~~\sigma_3 = \sqrt{\chi^2_0/\nu}\,\sigma_1,~~ \text{and} ~~~~\sigma_4 = \sqrt{2.3\chi^2_0/\nu}\,\sigma_1.
\end{equation}

Perhaps, the great variety of criteria in Table 1 are due to the fact that in most cases the standard rule $\chi^2 \leq \chi^2_0+1$ produces  too small uncertainties. This turns to be evident when the optimum values of the parameters are in effect changed by their uncertainties: the corresponding change in the angular distributions is often indistinguishable from the optimum ones.

\begin{table}
\caption{\label{tablacriterios} Different criteria used in the calculation of parameter uncertainties in the study of the threshold anomaly.
}
\begin{center}
\begin{tabular}{lclccll}
\br
 \multicolumn{2}{c}{Reference} & criterium  & Projectile & Target & $\langle \chi^2/N \rangle$  & conclusion \\
\mr
   {Biswas 2008  }	&	   \cite{bis08}	&	  { $\chi_0^2+ \chi_0^2/N$ }	 &	$^6$Li 	&	 $^{64}$Ni 	&	 0.43 	 &	 No TA \\[3pt]
	&		&		&	$^6$Li 	&	 $^{58}$Ni 	&	 0.45 	 &	 No conclusive \\[3pt]
   {Deshmuk 2011  }	&	   \cite{des11}	&	    {$\chi_0^2 + 1$ }	&	 $^6$Li 	&	 $^{112}$Sn 	&	 5.18 	 &	 No TA \\[3pt]
	&		&		&	$^6$Li 	&	 $^{116}$Sn 	&	 12.2 	&	 No TA \\[3pt]
 Fern\'andez Niello 2007  	&	 \cite{fer07} 	&	  $\chi_0^2 + 1$ 	&	 $^6$Li 	&	 $^{27}$Al 	&	 No data  	 &	 No TA  \\[3pt]
 Figueira  2006  	&	 \cite{fig06} 	&	 $\chi_0^2 + 1$ 	 &	$^7$Li 	 &	 $^{27}$Al 	 &	 3.6 	&	 No TA \\[3pt]
 {Figueira  2010  }	& \cite{fig10}	& $\chi_0^2 + 1$ 	&	 $^6$Li 	&	 $^{144}$Sm & 2.51 	& BTA \\[3pt]
	&		&		&	$^7$Li 	&	 $^{144}$Sm 	&	 2.68 	&	 No TA \\[3pt]
 Figueira 2007   	&	 \cite{fig07} 	&	  $\chi_0^2 + 1$ 	&	$^6$Li 	 &	 $^{27}$Al 	 &	 5.22 	&	 No TA \\[3pt]
     {Fimiani  2012 } 	&	     \cite{fim12}	&	     {$\chi_0^2 + 1$ }	 &	$^6$Li 	&	 $^{80}$Se 	&	 1.05 	 &	 BTA \\[3pt]
	&		&		&	$^7$Li 	&	 $^{80}$Se 	&	 1.00 	 &	 TA \\[3pt]
 Garc\'ia  2007   	&	 \cite{gar07}	&	 $\chi_0^2+N+1$ 	 &	$^6$He 	 &	 $^{209}$Bi  	 &	 2.2 	&	 BTA \\[3pt]
 Gomes 2004  	&	 \cite{gom04} 	&	  $\chi_0^2 + 1$ 	 &	$^9$Be 	&	 $^{27}$Al 	&	 No data 	&	  No TA \\[3pt]
 Gomes 2005  	&	 \cite{gom05} 	&	   $\chi_0^2+ \chi_0^2/N$   	&	 $^9$Be 	&	 $^{64}$Zn 	&	 No data 	 &	 BTA \\[3pt]
 G\'omez Camacho 2007  	&	 \cite{gom07} 	&	 no uncertainty bars 	&	 $^9$Be 	&	 $^{64}$Zn 	&	 0.61 	 &	  BTA \\[3pt]
 G\'omez Camacho  2008  	&	 \cite{gom08}	&	 no uncertainty bars 	 &	$^9$Be 	&	 $^{144}$Sm 	&	 0.61 	 &	 No TA  \\[3pt]
												
     G\'omez Camacho 2010  &	\cite{gom10}	&	 $\chi_0^2 + N $	&	 $^6$Li 	&	 $^{58}$Ni 	&	 0.61 	 &	 BTA \\[3pt]
	&		&		&	$^7$Be 	&	 $^{58}$Ni 	&	 0.05 	 &	  BTA \\[3pt]
	&		&		&	$^8$B  	&	 $^{58}$Ni 	&	 0.51/0.19	&	 BTA  \\[3pt]
     {Keeley 1994  }	&	     \cite{kee94}	&	    {$1.15\chi_0^2$  }	 &	$^6$Li 	&	 $^{208}$Pb 	&	 2.16 	 &	 No TA \\[3pt]
	&		&		&	$^7$Li 	&	 $^{208}$Pb 	&	 1.53 	&	 weak TA \\[3pt]
 Kumawat 2008 	&	 \cite{kum08} 	&	 $1.5 \chi_0^2 $	 &	$^6$Li 	&	 $^{90}$Zr 	&	 0.82 	&	 No TA \\[3pt]
 Lubian 2003  	&	 \cite{lub03}	&	  $\chi_0^2 + 1$ 	 &	$^7$Li 	&	 $^{138}$Ba 	 &	 No data 	&	 BTA \\[3pt]
     {Maciel 1999  }	&	    \cite{mac99}	&	     {qualitative }	&	 $^6$Li 	&	 $^{138}$Ba 	&	 No data 	 &	 No TA \\[3pt]
	&		&		&	$^7$Li 	&	 $^{138}$Ba 	&	 No data 	&	 TA \\[3pt]
 Nagarajan 1985  	&	 \cite{nag85}	&	 no uncertainty bars 	&	 $^{16}$O 	&	 $^{208}$Pb 	&	 No data 	 &	 TA \\
 Pakou 2003   	&	 \cite{pak03}  	&	 Not given 	&	 $^6$Li 	&	 $^{28}$Si 	&	  No data 	&	 No TA \\[3pt]
 Pakou 2004  	&	 \cite{pak04} 	&	 Not given 	&	 $^7$Li 	&	 $^{28}$Si 	&	 No data 	&	 No TA \\[3pt]
 Santra 2001  	&	 \cite{san01}	&	$2\chi_0^2$ 	&	 $^{12}$C 	&	 $^{208}$Pb 	 &	 6.03 	&	 TA \\[3pt]
     {Souza 2007 } 	&	    { \cite{sou07} }	&	    { $1.1 \chi_0^2$     }	&	 $^6$Li 	 &	 $^{59}$Co 	&	 No data 	 &	 TA or BTA  \\[3pt]
	&		&		&	$^7$Li 	&	 $^{59}$Co 	&	 No data 	&	 TA \\[3pt]
 Woolliscroft  2004  	&	 \cite{woo04}	&	  $1.15\chi_0^2$  	&	 $^9$Be 	&	 $^{208}$Pb 	&	11	 &	 TA  \\[3pt]
 Zadro 2009  	&	 \cite{zad09}	&	  $\chi_0^2 + 1$ 	 &	$^6$Li 	&	 $^{64}$Zn 	&	 1.18 	&	 No TA \\[3pt]
\br
\end{tabular}
\end{center}
\end{table}

\subsection{Error ellipses}\label{errorellipses}
One interesting example within low-energy nuclear physics of the use of error ellipses to estimate uncertainties for correlated parameters  is presented in Ref.~\cite{bar05}. However, this has not been applied to take into account the correlation between the optical-model parameters. In fact, none of the works listed in Table \ref{tablacriterios} uses this technique.  In our case the ellipses corresponding to the different limits previously mentioned  are calculated as
\begin{equation}\label{ecuacionelipse}
\displaystyle{\Delta \chi^2{'} = (\boldsymbol{\alpha}-\boldsymbol{\alpha_0})^T C^{-1} (\boldsymbol{\alpha}-\boldsymbol{\alpha_0})},
\end{equation}
where $\Delta \chi^2{'}$ is defined in eq.\ref{chi2+chi20/nudeltachi}, the covariance matrix $C$ is calculated for each system and energy and $\boldsymbol{\alpha}-\boldsymbol{\alpha_0}$ represent deviations respect the optimum value of the parameters.

\begin{figure}
\begin{center}
\includegraphics[trim=0cm 0cm 0cm .8cm, clip=true, width=36pc]{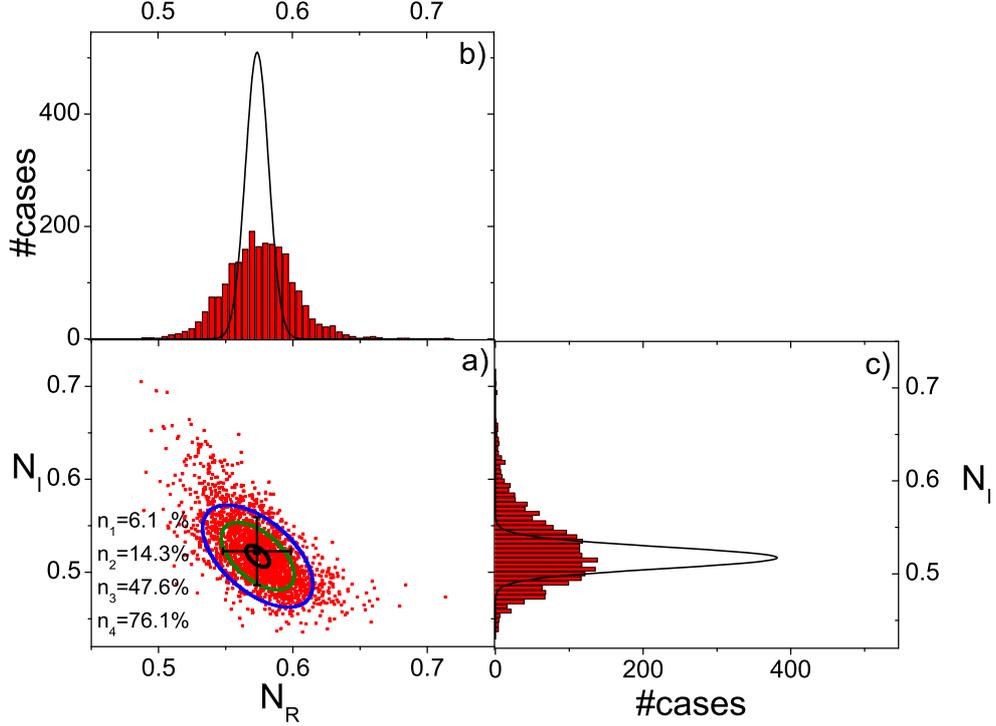}
\end{center}
\vspace{-1cm}
\caption{\label{figuraelipses} Plot a) shows the bootstrapped values (red points) and the ellipses described in section \ref{errorellipses} ($\sigma_2$ ellipses not shown for clarity). Histograms in  b) and c) are the projection of bootstrapped points on the $N_R$ and $N_I$ axis respectively, while the full line shows the much narrower gaussian distribution with standard deviation $\sigma_1$ ($\chi\leq\chi_0^2+1$) for comparison. }
\end{figure}

Figure \ref{figuraelipses} shows three of such ellipses for the $^{12}$C + $^{208}$Pb angular distribution at 84.9~MeV corresponding to the $\sigma_1$, $\sigma_3$ and $\sigma_4$ limits for the determination of the standard uncertainty in both parameters. The limit $\sigma_2$ should be used in cases where $(\chi^2/\nu) \approx \ 1$. In the cases were $(\chi^2/\nu) > 1$, $\sigma_4$ should be used. We will show the error ellipses for all energies of both cases studied in Figs. \ref{elipses12c208pb} and \ref{elipses6li80Se}.

\subsection{Bootstrap}\label{bootstrap}

Bootstrap is one of the many resampling methods designed to go beyond regular statistic test. It creates a number $N_B$ of synthetic data sets, each of them consisting on $N$ data points selected randomly (with reposition) from the original data set of $N$ points  (it is desirable to have  $N_B \gg N$). On the average a fraction $1/e$ (about  37$\%$) of the $N$ points will be repeated ones (thus having more weight in the fitting procedure) and, consequently, 37$\%$ of the elements will not be included. Each synthetic data set is used to fit  the parameters of interest. The standard deviation of the $N_B$  values obtained for each parameter is a reliable estimation of its uncertainty. Hence, this procedure gauges the sensitivity of the fitted parameter to each individual data point by the simulation of $N_B$ experiments in which the experimenter could have chosen to repeat the measurement of some of the data points at the cost of missing some others. The bootstrap method has been one of the techniques applied to the evaluation of the half-life of $^{198}$Au in Ref.~\cite{che11}. To the best of our knowledge, bootstrap techniques have not been applied yet to the calculation of nuclear potential parameters and their uncertainties.

In Fig. \ref{figuraelipses} we show the resulting $N_R, N_I$ points from the adjustment of $N_B=2200$ synthetic angular distributions. The average and variance converged for $N_B \geq 200$. It is interesting to point out that 76$\%$ of the bootstrap points  are enclosed by the $\sigma_4$ ellipse.
\section{Case studies}\label{SectionCaseStudies}
In the first step to study the  $^{12}$C + $^{208}$Pb and the $^6$Li + $^{80}$Se systems, the optimum $N_R$, $N_I$ values are calculated as well as the corresponding residuals, covariance matrices and error ellipses. Since a condition of the analysis is the normality of the residuals, the Kolmogorov-Smirnov test  has been applied. This statistical test evaluates the maximum difference between the empirical and reference (theoretical) cumulative distribution functions and yields the acceptance or rejection of a null hypothesis at a given confidence level \cite{fro79}.

In our case, we compare to two different null hypothesis: \emph{i)} the residuals follows a normal distribution with mean value $\langle R \rangle = 0$  and variance $\sigma^2_R = 1$, i.e. $H_{01}:R_{i}\sim {\cal N}(0,1)$ and  \emph{ii)} the residuals follows a normal distribution but in this case with variance $\sigma^2_R = \chi_0^{2}/\nu $ i.e. $H_{02}:R_{i}\sim {\cal N}(0,\chi_0^{2}/\nu)$.


\subsection{The $^{12}$C + $^{208}$Pb system}\label{12Csubsect}

\begin{figure}
\begin{center}
\includegraphics[trim=0cm 0.1cm 0cm .5cm, clip=true,width=40pc]{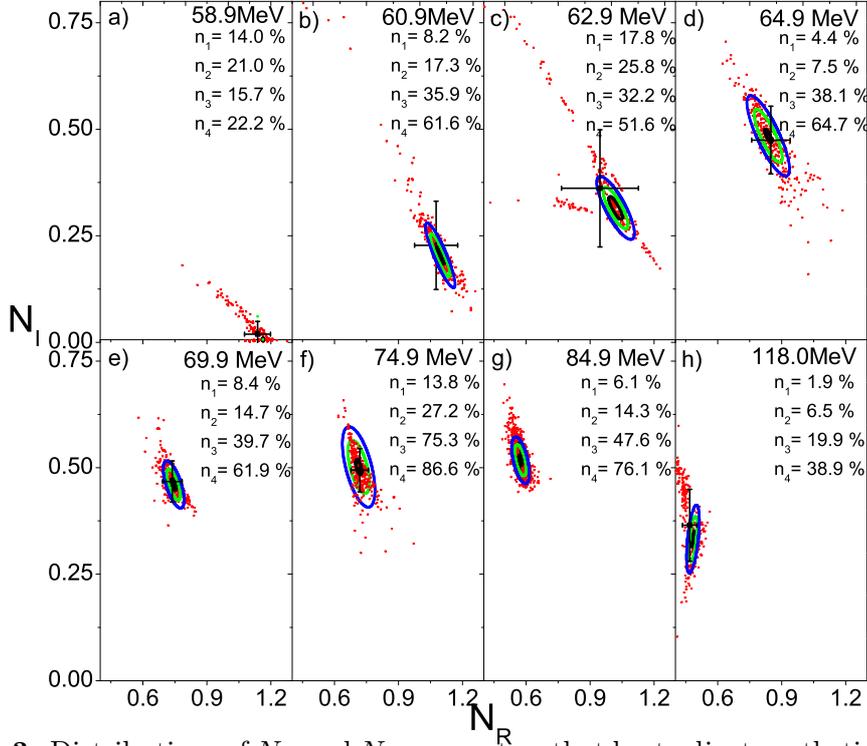}
\end{center}
\vspace{-2.5cm}
\caption{\label{elipses12c208pb}Distributions of $N_R$ and $N_I$ parameters that best adjust synthetic data sets generated through the bootstrap technique for the $^{12}$C + $^{208}$Pb system. The analysis based on the ellipses is described in section 3.3. The uncertainty bars indicate the medium values of the parameter distribution and their standard deviations.}
\end{figure}

\begin{figure}
\begin{center}
\includegraphics[trim=0cm 0cm 0cm 0cm, clip=true,width=32pc]{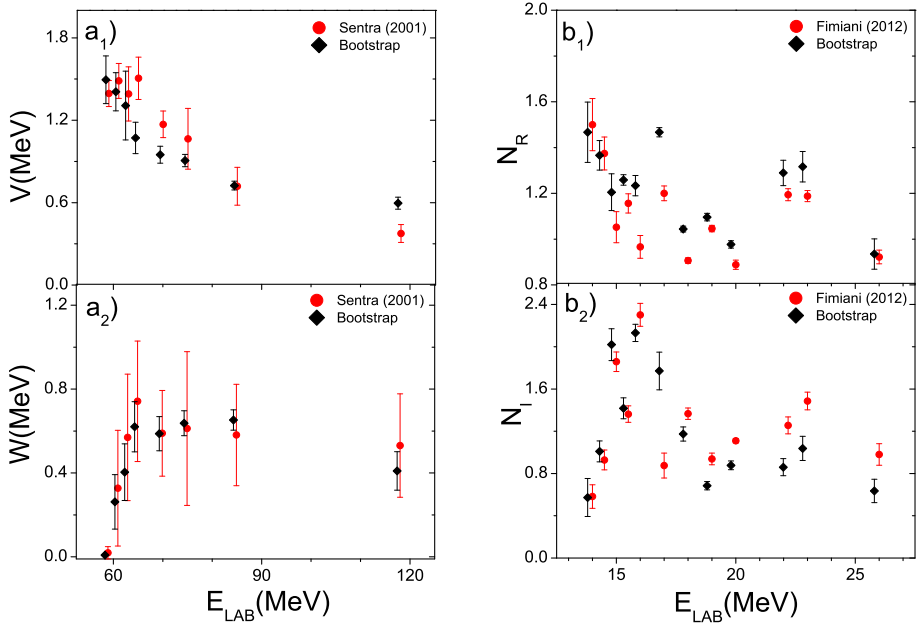}
\end{center}
\vspace{-1.9cm}
\caption{\label{DR}Comparison between the dispersion relations reported in the original works and those obtained in the present analysis with the bootstrap technique a) for the $^{12}$C + $^{208}$Pb system and b) for the $^6$Li + $^{80}$Se system. The bootstrap results are slightly displaced in energy for clarity.}
\end{figure}

As expected, the residuals of the angular distributions of each individual energy, as well as the residuals corresponding to all data taken together, follow ${\cal N}(0,\chi_0^{2}/\nu)$ but do not follow ${\cal N}(0,1)$. Fig. \ref{elipses12c208pb} shows the error ellipses corresponding to the limits $\sigma_1$, $\sigma_3$ and $\sigma_4$ of eq. \ref{criteriossigma} ($\sigma_2$ not shown for clarity) and points ($N_R$, $N_I$) representing the results from fitting bootstrapped data sets as explained in section \ref{bootstrap}. For all energies, $N_B \geq 300$. The numbers $n_1$, $n_2$, $n_3$ and $n_4$ indicate the number of these bootstrapped points inside each ellipse (as it has been shown in Fig. \ref{figuraelipses}, expressed as a percentage of $N_B$). Except for the lowest and highest energies more than 50$\%$ of the bootstrapped points are enclosed inside the $\sigma_4$ ellipse. In general  they extend beyond  the last ellipse but keeping the same correlation. If cases c) and h) of this figure a group of points diverge from the main trend. The effect was traced to the influence of one single experimental point in both cases, been the divergent group part of a population in which that point is omitted. In Fig. \ref{DR}, the parameters and their uncertainties given by the bootstrap method  is compared with the ones given by Ref.\cite{san01}. It is seen that even though our assigned uncertainties are smaller the character of a regular TA is maintained.

\subsection{$^6$Li + $^{80}$Se system}\label{6Lisubsect}
\begin{figure}
\begin{center}
\includegraphics[trim=0.8cm 0cm 0cm .6cm, clip=true,width=40pc]{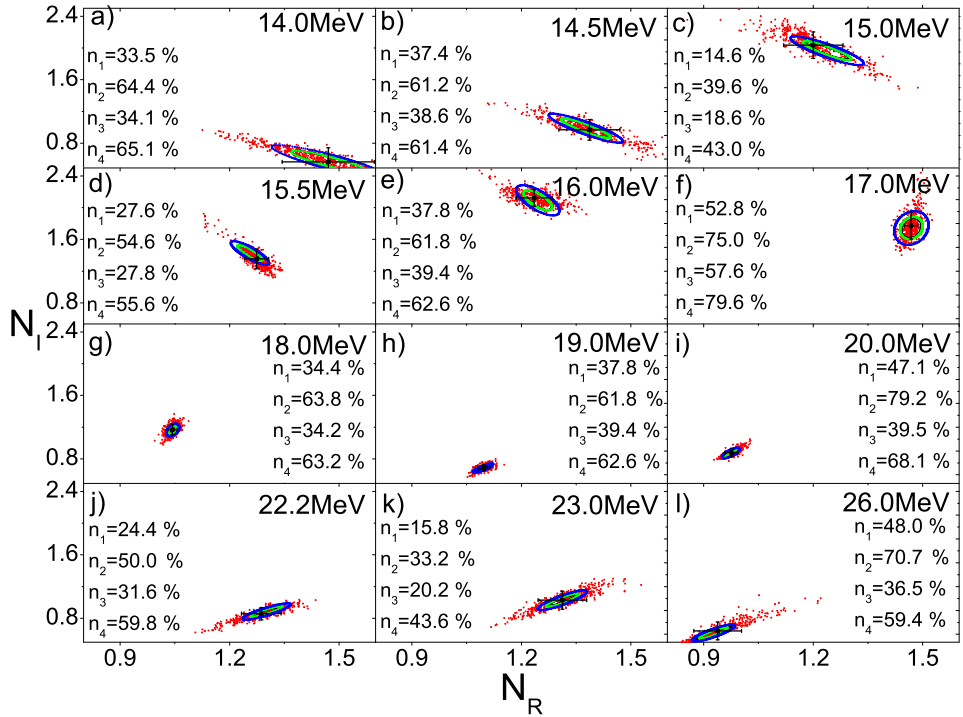}
\end{center}
\vspace{-4.0cm}
\caption{\label{elipses6li80Se}Distributions of $N_R$ and $N_I$ parameters that best adjust synthetic data sets generated by the bootstrap technique for the $^6$Li + $^{80}$Se system. The analysis based on the ellipses is described in section 3.3. The uncertainty bars indicate the medium values of the parameter distribution and their standard deviations.}
\end{figure}

  Since in this experiment $\chi_0^2/\nu \approx  1$  both null hypothesis almost coincide. The Kolmogorov-Smironov test  indicates that  residuals corresponding to all data taken together are normally distributed, and so are individual angular distributions for each energy, except for the data corresponding to 23 and 14.5 MeV. This reveals that for these two energies there is a subtle difference between experimental data and the theoretical model. Even though this difference is statistically significant, it would have remained unnoticed without this analysis. Normalization factors  of around $1.6\%$ applied to these angular distributions are enough to make them pass the test, but this may require further analysis.  Figure \ref{elipses6li80Se} shows the error ellipses and bootstrapped points as in Fig. \ref{figuraelipses}. Here, more than 43$\%$ of the bootstraped points are enclosed inside the $\sigma_4$ ellipse.
 The behavior of the parameters and their uncertainties given by the bootstrap limits is compared with the ones given by in Ref.~\cite{fim12} in Fig. \ref{DR}.

\vspace{-2cm}

\begin{nopagebreak}[4]

\vspace{-2cm}
\section{Conclusions}\label{SectionConclusions}
 We have shown that most studies use the standard rule from curve fitting: $\chi^2 \leq  \chi^2_{0}+1$  (eq. \ref{chi2+1}), to determine the uncertainty in fitting parameters (usually $N_R$ and $N_I$) from experimental data, regardless of their experimental value of $\chi_0^2/\nu$. This is not satisfactory even from the point of view of conventional statistic:
 For the cases where $\chi_0^2/\nu > 1$ that prescription is incorrect and should be replaced by  $\chi^2 \leq \chi^2_0+2.3\chi^2_0/\nu$ of eq. \ref{sigma4}. Parameters uncertainties which has been already derived using the $\chi^2 \leq  \chi^2_{0}+1$   recipe , can be easily scaled up (under the parabolic approximation of eq. \ref{receta})  by a factor of $\sqrt{2.3\chi^2_0/\nu}$.

The bootstrap resampling method applied in this work produced a more realistic distribution of the parameters as shown in Figs. \ref{figuraelipses} b) and c). The resulting uncertainties  are similar or somewhat larger than the ones obtained with the $\sigma_4$ ellipse (eq.  \ref{sigma4}).

To judge the kind of TA it is very helpful to look at the series of 2-dim  $N_R$ vs. $N_I$ plots with their respective uncertainty ellipses. In this way, the correlation between the parameters can be considered. In Fig. \ref{DR} our results, plotted in the conventional way, show a more reliable estimation of the parameter uncertainties than the quoted in the original works (smaller for $^{12}$C + $^{208}$Pb, similar or slightly larger for  $^6$Li + $^{80}$Se). In these cases, however, the conclusions about the kind of TA was not modified. Further work is in progress to extend the present analysis to other systems, using bootstrap as well as other resampling techniques.

\end{nopagebreak}
\end{document}